\documentclass{article}
\usepackage{amssymb}
\usepackage{tikz}
\usetikzlibrary{arrows, calc, shapes, patterns, spy, backgrounds, fit, matrix, positioning}
\usepackage{epstopdf}
\newcommand{\phos}{Pi}
\newcommand{\atp}{ATP}
\newcommand{\dhap}{DHAP}

\newcommand{\gap}{GAP}

\newcommand{\dpga}{DPGA}
\newcommand{\pga}{PGA}

\newcommand{\nadp}{NADP}
\newcommand{\nadph}{NADPH}
\newcommand{\adp}{ADP}
\newcommand{\kohlenstoffdioxid}{{CO_2}}
\newcommand{\fbp}{FBP}
\newcommand{\rubp}{RuBP}
\newcommand{\glp}{G6P}
\newcommand{\fp}{F6P}
\newcommand{\glup}{G1P}
\newcommand{\rup}{Ru5P}
\newcommand{\xp}{X5P}
\newcommand{\ep}{E4P}
\newcommand{\sbp}{SBP}
\newcommand{\ssp}{S7P}
\newcommand{\rp}{R5P}
\newcommand{\st}{Starch}

\begin{document}

\title{Overload breakdown in models for photosynthesis} 

\author{Dorothea M\"ohring and Alan D. Rendall\\
Institut f\"ur Mathematik\\
Johannes Gutenberg-Universit\"at\\
Staudingerweg 9\\
D-55099 Mainz\\
Germany}

\date{}

\maketitle

\begin{abstract}
In many models of the Calvin cycle of photosynthesis it is observed that
there are solutions where concentrations of key substances belonging to the
cycle tend to zero at late times, a phenomenon known as overload breakdown.
In this paper we prove theorems about the existence and non-existence of
solutions of this type and obtain information on which concentrations tend to
zero when overload breakdown occurs. As a starting point we take a model of
Pettersson and Ryde-Pettersson which seems to be prone to overload 
breakdown and a modification of it due to Poolman which was intended to
avoid this effect.
\end{abstract}

\section{Introduction}

Photosynthesis is one of the most important processes in biology and a variety
of mathematical models have been set up in order to describe it. These have
mainly been concerned with the part of photosynthesis known as the dark
reactions, also known as carbon fixation or the Calvin cycle. In the simplest
pictures of this process (see e.g. \cite{alberts02}) the substances included
in the description are the five carbohydrate phosphates RuBP, PGA, DPGA, GAP 
and Ru5P. In \cite{grimbs11} the properties of some models of this type were 
considered, together with a model which in addition includes the concentration 
of ATP as a variable. In \cite{rendall14} dynamical properties of solutions of 
these models and related ones were studied. It was found that for many of
the models there are solutions where the concentrations become unboundedly 
large at late times although they remain finite on all finite time intervals
(runaway solutions). The one exception is the model including ATP, for which 
it was shown that all solutions are globally bounded. At the same time it was 
shown for all the models considered that there are large classes of initial 
data for which the corresponding solutions are such that the concentrations
of all carbohydrate phosphates tend to zero as time tends to infinity. In other 
words all these concentrations become arbitrarily small at late 
times. In \cite{rendall14} no biological interpretation was offered for this 
behaviour. Both these phenomena, where the concentrations become arbitrarily 
large or small, might be taken as indications that these models are 
inappropriate. It suggests that it would be worthwhile to examine alternative 
models to see if they exhibit similar behaviour.

In this paper we study some models of the Calvin cycle which incorporate more
aspects of the biology. In the simple models the five substances included
are linked by reactions forming a cycle. In the models considered in what 
follows this simple circular topology of the reaction network is replaced 
by a more complicated branched one. The starting point is a model 
introduced by Pettersson and Ryde-Pettersson in \cite{pettersson88}. We
call this the Pettersson model. The unknowns are concentrations of substances
in the chloroplast, where the Calvin cycle takes place. Most of the reactions
in the model convert some of these substances into others. It also includes some
transport processes, where substances are exported from the chloroplast to
the surrounding cytosol and a process in which they are stored as starch in the
chloroplast. It was found in \cite{pettersson88} that for certain parameter
values, corresponding to high concentrations of inorganic phosphate in the 
cytosol, solutions of this model can undergo a process called overload breakdown
where the export of sugars cannot be maintained. Since this behaviour may 
be biologically unrealistic Poolman \cite{poolman99} introduced a new model, 
which we call the Poolman model, in an attempt to avoid solutions of this
type. Overload breakdown corresponds to a situation where more sugar phosphates
are being exported from the system than can be produced and to combat this
Poolman introduced an extra reaction describing starch degradation, i.e. 
the production of sugar phosphates from starch. This is the only difference
between the reaction networks used in the Pettersson and Poolman models. There 
is also a difference in the kinetics which describe the mechanisms of certain
reactions. Yet another choice of kinetics, which has the advantage of 
simplicity, is mass action kinetics. The models obtained from the two networks
by applying mass action kinetics will be called the Pettersson-MA and 
Poolman-MA models. For all these models it is easy to see that all solutions
remain bounded, due to the conservation of the total amount of phosphate. 
They have no runaway solutions. In what follows we investigate under what 
circumstances these models with mass action kinetics admit solutions where 
some concentrations tend to zero at late times and to what extent these 
conclusions can be transferred to the original Pettersson and Poolman models.

The paper proceeds as follows. In Sect. 2 the various models are introduced
and the relations between them are described. The question, which combinations 
of the variables in the system might tend to zero at late times is examined in 
Sect. 3. In other words, necessary conditions for a point on the boundary
of the state space to be an  $\omega$-limit point of a positive solution are
obtained. It is shown in Sect. 4 that in the case of the Pettersson-MA model
there are large classes of solutions for which many concentrations converge to 
zero as $t\to\infty$ and it is investigated to what extent this is prevented 
by moving to the Poolman-MA model. The set of substances whose concentrations
may tend to zero gives a new picture of the details of overload breakdown. 
The main results on the Pettersson-MA and Poolman-MA models are contained in 
Theorem 1. Sect. 5 is concerned with generalizing some of these results to 
the original Poolman model. Conclusions and an outlook are given in the last 
section.

This paper is based in part on the master's thesis of the first author
\cite{moehring15}.
 
\section{Basic definitions}

This section gives the definitions of four models of the Calvin cycle
mentioned in the introduction, the Pettersson-MA, Poolman-MA, Poolman and
Pettersson models. The underlying reaction network is given in Figure 1.

\begin{figure}[h]
\begin{center}
	\begin{tikzpicture}[scale=0.80]
		\coordinate (v1) at (0.9, 1.2) {};
	% --- Punkte von Ebene 0 bis 13 ------------------
	\begin{scope}[every node/.style={font=\sffamily\small}]
		\node[] (PiA) at (0,2) {\phantom{ext}$\phos$};
		\node[] (PiextA) at (0,1) {$\phos_{ext}$};
		\node[] (ATPC) at (-2,3) {$\atp$};
		\node[] (DHAP) at (1,3) {$\dhap$};
		\node[] (DHAPext) at (1,0) {$\dhap_{ext}$};
		\node[] (PiB) at (3,2) {$\phos$};
		\node[] (PiextB) at (3,1) {$\phos_{ext}$};
		\node[] (GAP) at (4,3) {$\gap$};
		\node[] (GAPext) at (4,0) {$\gap_{ext}$};
		\node[] (DPGA) at (7,3) {$\dpga$};
		\node[] (PiC) at (9,2) {$\phos$};
		\node[] (PiextC) at (9,1) {$\phos_{ext}$};
		\node[] (PGA) at (10,3) {$\pga$};
		\node[] (PGAext) at (10,0) {$\pga_{ext}$};
		\node[] (PiD) at (5,2.3) {$\phos$};
		\node[] (NADP) at (4.9,4) {$\nadp$};
		\node[] (NADPH) at (6.3,4) {$\nadph$};
		\node[] (ADPA) at (8,4) {$\adp$};
		\node[] (ATPA) at (9,4) {$\atp$};
		\node[] (PiE) at (-1, 5) {$\phos$};
		\node[] (CO) at (11,5) {$\kohlenstoffdioxid$};
		\node[] (ADPC) at (-2,6) {$\adp$};
		\node[] (FBP) at (2.5,6) {$\fbp$};
		\node[] (RuBP) at (10,6) {$\rubp$};
		\node[] (PiF) at (3.5,7) {$\phos$};
		\node[] (ADPB) at (9,7) {$\adp$};
		\node[] (GlP) at (-2,8) {$\glp$};
		\node[] (FP) at (2.5,8) {$\fp$};
		\node[] (ATPB) at (9,8) {$\atp$};
		\node[] (GluP) at (-2,9) {$\glup$};
		\node[] (RuP) at (10,9) {$\rup$};
		\node[] (ATPD) at (-0.5,10) {$\atp$};
		\node[] (XP) at (8,10) {$\xp$};
		\node[] (ADPD) at (-0.5,11) {$\adp$};
		\node[] (EP) at (2.5,11) {$\ep$};
		\node[] (PiG) at (-3.5,12) {$\phos$};
		\node[] (PiH) at (-0.5,12) {$\phos$};
		\node[] (PiJ) at (5.3,12) {$\phos$};
		\node[draw,ellipse] (starch) at (-2,13) {$\st$};
		\node[] (SBP) at (4,13) {$\sbp$};
		\node[] (SP) at (6,13) {$\ssp$};
		\node[] (RP) at (10,13) {$\rp$};
	% --- Stroma und Cytosol ---------------------------
		\node[] (Stroma) at (-2.8,2) {Stroma};
		\node[] (Stroma) at (-2.8,1) {Cytosol};
	% --- gebogene Pfeile -------------------------------------------------------------------------------------
		\draw[->, thick] (PiextA) to [out=0, in=-90](1, 1.3) to [out=90, in=90](1, 1.6) to [out=90, in=0](PiA);
		\draw[->, thick] (PiextB) to [out=0, in=-90](4, 1.3) to [out=90, in=90](4, 1.6) to [out=90, in=0](PiB);
		\draw[->, thick] (PiextC) to [out=0, in=-90](10, 1.3) to [out=90, in=90](10, 1.6) to [out=90, in=0](PiC);
		\draw[->, thick] (DPGA) to [out=180, in=0](5.4,3) to [out=180, in=90](PiD);
		\draw[<->, thick] (NADP) to [out=-90, in=180](5.4,3) to [out=0, in=180](5.7,3) to [out=0, in=-90](NADPH);
		\draw[<->, thick] (ADPA) to [out=-90, in=180](8.4,3) to [out=0, in=180](8.6,3) to [out=0, in=-90](ATPA);
		\draw[->, thick] (PiE) to [out=180, in=90](-2,4) to [out=-90, in=90](ATPC);
		\draw[<->, thick] (FBP) to [out=-90, in=90](2.5,4.5) to [out=-90, in=135](GAP);
		\draw[<->, thick] (FBP) to [out=-90, in=90]node[left]{$k_{8,9}$}(2.5,4.5) to [out=-90, in=45](DHAP);
		\draw[->, thick] (CO) to [out=180, in=90](10,4.5) to [out=-90, in=90] node [right, near start] {$k_1$} (PGA);
		\draw[->, thick] (FBP) to [out=90, in=-90](2.5,6.5) to [out=90, in=180](PiF);
		\draw[->, thick] (ATPB) to [out=0, in=90](10,7.6) to [out=-90, in=90](10,7.4) to [out=-90, in=0](ADPB);
		\draw[->, thick,dotted] (starch) to [out=-120, in=90](-2.6,11) 
																										to [out=-90, in=120] node [left] {$k_{32}$} (GluP);
		\draw[-, thick,dotted] (PiG) to [out=0, in=90](-2.56,11.6);
		\draw[->, thick] (GluP) to [out=70, in=-90](-1.5,11) to [out=90, in=-70]node [right, near end] {$k_{27}$} (starch);
		\draw[->, thick] (-1.56,11.6) to [out=90, in=180](PiH);
		\draw[->, thick] (ATPD) to [out=180, in=-90](-1.54,10.4) to [out=90, in=-90](-1.52, 10.6) to [out=90, in=180](ADPD);
		\draw[<->, thick] (DHAP) to [out=90, in=-90](1,12.6) to [out=90, in=180](1.4,13) 
														 to [out=0, in=180] node [above] {$k_{13,14}$} (SBP);
		\draw[<->, thick] (EP) to [out=90, in=-90](2.5,12.6) to [out=90, in=180](2.9,13) to [out=0, in=180](SBP);
		\draw[->, thick] (SBP) to [out=0, in=180](4.8,13) to [out=0, in=90](PiJ);
		\draw[<->, thick] (XP) to [out=180, in=0](3,10) to [out=180, in=90](2.5,9.5) to [out=-90, in=180](3,9) 
														to [out=0, in=180](3.6,9) to [out=0, in=90](4,8.6) to [out=-90, in=90](GAP);
		\draw[<->, thick] (XP) to [out=90, in=-90](8,12.6) to [out=90, in=0](7.5,13) to [out=180, in=90](7,12.6)
														to [out=-90, in=90](7,6.4) to [out=-90 ,in=0](6.4,6) to [out=180,in=0](4.4,6)
														to [out=180,in=90](4,5.6) to [out=-90, in=90](GAP);
	% --- Rechtecke -------------------------------------------------------
		\draw[pattern = vertical lines] (0.9, 1.3) rectangle +(-4, 0.4);
		\draw[pattern = vertical lines] (3.9, 1.3) rectangle +(-2.8, 0.4);
		\draw[pattern = vertical lines] (9.9, 1.3) rectangle +(-5.8, 0.4);
		\draw[pattern = vertical lines] (10.1, 1.3) rectangle +(1.25, 0.4);
		%\draw[pattern = north east lines] (10.1, 1.3) rectangle +(1.25, 0.4); %werden in pdf nicht schön umgestzt
		\draw[dotted, rounded corners] (-0.5,2.5) rectangle +(-2.5,4);
		\end{scope}
	% --- Direkte Pfeile -----------------------------------------------------
		\path[every node/.style={font=\sffamily\small}]
		(DHAP) edge [->, thick] node [right, very near start] {$k_{30}$} (DHAPext)
		(GAP) edge[->, thick] node [right, very near start] {$k_{29}$} (GAPext)
		(PGA) edge[->, thick] node [right, very near start] {$k_{28}$} (PGAext)
		(DHAP) edge [<->, thick] node [below] {$k_{6,7}$} (GAP)
		(GAP) edge [<->, thick] node [below, near end] {$k_{4,5}$} (DPGA)
		(DPGA) edge [<->, thick] node [below] {$k_{2,3}$} (PGA)
		(ADPC) edge [->, thick] node [left] {$k_{31}$} (ATPC)
		(RuBP) edge [->, thick] (PGA)
		(FBP) edge [->, thick] node [left] {$k_{10}$} (FP)
		(RuP) edge [->, thick] node [right] {$k_{22}$} (RuBP)
		(GlP) edge [<->, thick] node [below] {$k_{23,24}$} (FP)
		(GluP) edge [<->, thick] node [right] {$k_{25,26}$} (GlP)
		(EP) edge [<->, thick] node [left] {$k_{11,12}$} (FP)
		(SBP) edge [->, thick] node [above] {$k_{15}$} (SP)
		(SP) edge [<->, thick] node [above] {$k_{16,17}$} (RP)
		(RP) edge [<->, thick] node [right] {$k_{18,19}$} (RuP)
		(XP) edge [<->, thick] node [above, near end] {$k_{20,21}$} (RuP);
		\end{tikzpicture}
	\end{center}
	\caption{diagram of the network}
	\label{fig:diagram}
\end{figure}
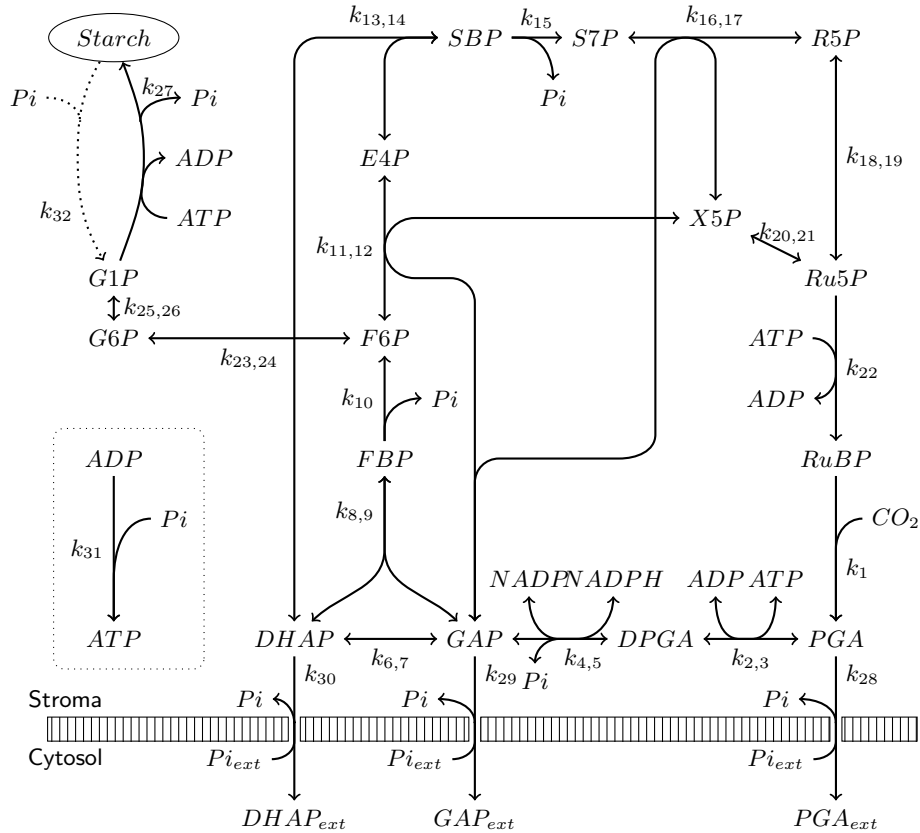

%\begin{figure}
%\includegraphics{diagram.eps}
%\caption{The network describing the Calvin cycle}
%\end{figure}

\vskip 10pt\noindent
In the case of the Pettersson model $k_{32}=0$ and one of the reactions can 
be dropped. In the case of the Poolman model $k_{32}\ne 0$. The system of
ordinary differential equations obtained by applying mass action kinetics 
in all reactions will now be presented. To our knowledge this model has not
previously appeared explicitly in the literature although it does occur 
implicitly in \cite{grimbs99}. In that reference the author investigates the 
applicability of some theorems of chemical reaction network theory to this 
system and since the hypotheses of these theorems include mass action kinetics 
the only logical interpretation is that he applied these techniques to the mass
action system arising from the given network. The unknowns in the system
are the concentrations of RuBP, PGA, DPGA, ATP, GAP, $\rm P_i$, DHAP, FBP, 
F6P, E4P, X5P, SBP, S7P, R5P, Ru5P, G6P and G1P. The equations are
\begin{eqnarray}
&&\frac{d x_{RuBP}}{dt}=-k_1 x_{RuBP}+k_{22} x_{Ru5P} x_{ATP},\\
&&\frac{d x_{PGA}}{dt}=2k_1 x_{RuBP}+k_2 x_{DPGA}(c_A-x_{ATP})\nonumber\\
&&-k_3x_{PGA}x_{ATP}
-k_{28}x_{PGA},\\
&&\frac{d x_{DPGA}}{dt}=-k_2 x_{DPGA}(c_A-x_{ATP})+k_3x_{PGA}x_{ATP}\nonumber\\
&&+k_4x_{GAP}x_{P_i}-k_5x_{DPGA},\\
&&\frac{d x_{ATP}}{dt}=k_2 x_{DPGA}(c_A-x_{ATP})-k_3x_{PGA}x_{ATP}-k_{22} x_{Ru5P} x_{ATP}
\nonumber\\
&&-k_{27}x_{G1P}x_{ATP}+k_{31}x_{P_i}(c_A-x_{ATP}),\\
&&\frac{d x_{GAP}}{dt}=-k_4x_{GAP}x_{P_i}+k_5x_{DPGA}+k_6x_{DHAP}-k_7x_{GAP}+k_8x_{FBP}
\nonumber\\
&&-k_9x_{GAP}x_{DHAP}+k_{11}x_{E4P}x_{X5P}-k_{12}x_{F6P}x_{GAP}+k_{16}x_{X5P}x_{R5P}
\nonumber\\
&&-k_{17}x_{S7P}x_{GAP}-k_{29}x_{GAP},\\
&&\frac{d x_{P_i}}{dt}=-k_4x_{GAP}x_{P_i}+k_5x_{DPGA}+k_{10}x_{FBP}+k_{15}x_{SBP}
\nonumber\\
&&+2k_{27}x_{G1P}x_{ATP}+k_{28}x_{PGA}+k_{29}x_{GAP}+k_{30}x_{DHAP}\nonumber\\
&&-k_{31}x_{P_i}(c_A-x_{ATP})-k_{32}x_{P_i},\\
&&\frac{d x_{DHAP}}{dt}=-k_6x_{DHAP}+k_7x_{GAP}+k_8x_{FBP}-k_9x_{GAP}x_{DHAP}\nonumber\\
&&+k_{13}x_{SBP}-k_{14}x_{DHAP}x_{E4P}-k_{30}x_{DHAP},\\
&&\frac{d x_{FBP}}{dt}=-k_8 x_{FBP}+k_9x_{GAP}x_{DHAP}-k_{10}x_{FBP},\\
&&\frac{d x_{F6P}}{dt}=k_{10}x_{FBP}+k_{11}x_{E4P}x_{X5P}-k_{12}x_{F6P}x_{GAP}\nonumber\\
&&+k_{23}x_{G6P}-k_{24}x_{F6P},\\
&&\frac{d x_{E4P}}{dt}=-k_{11}x_{E4P}x_{X5P}+k_{12}x_{F6P}x_{GAP}
\nonumber\\
&&+k_{13}x_{SBP}-k_{14}x_{DHAP}x_{E4P},\\
&&\frac{d x_{X5P}}{dt}=-k_{11}x_{E4P}x_{X5P}+k_{12}x_{F6P}x_{GAP}-k_{16}x_{X5P}x_{R5P}
\nonumber\\
&&+k_{17}x_{S7P}x_{GAP}+k_{20}x_{Ru5P}-k_{21}x_{X5P},\\
&&\frac{d x_{SBP}}{dt}=-k_{13}x_{SBP}+k_{14}x_{DHAP}x_{E4P}-k_{15}x_{SBP},\\
&&\frac{d x_{S7P}}{dt}=k_{15}x_{SBP}+k_{16}x_{X5P}x_{R5P}-k_{17}x_{S7P}x_{GAP},\\
&&\frac{d x_{R5P}}{dt}=-k_{16}x_{X5P}x_{R5P}+k_{17}x_{S7P}x_{GAP}+k_{18}x_{Ru5P}
-k_{19}x_{R5P},\\
&&\frac{d x_{Ru5P}}{dt}=-k_{18}x_{Ru5P}+k_{19}x_{R5P}-k_{20}x_{Ru5P}\nonumber\\
&&+k_{21}x_{X5P}-k_{22}x_{Ru5P} x_{ATP},\\
&&\frac{d x_{G6P}}{dt}=-k_{23}x_{G6P}+k_{24}x_{F6P}+k_{25}x_{G1P}-k_{26}x_{G6P},\\
&&\frac{d x_{G1P}}{dt}=-k_{25}x_{G1P}+k_{26}x_{G6P}-k_{27}x_{G1P}x_{ATP}
+k_{32}x_{P_i}
\end{eqnarray}
where $x_X$ denotes the concentration of any substance $X$ and the $k_i$ are 
constants, the reaction constants. These are all positive with the possible 
exception of $k_{32}$. When it is zero the equations define the Pettersson-MA 
model and when it is positive they define the Poolman-MA model. 
$c_A=x_{ADP}+x_{ATP}$ is the total concentration of adenosine phosphates. It
is constant in time and this fact has been used to eliminate $x_{ADP}$ from
the evolution equations. The concentrations of substances in the cytosol
in the diagram of the network, which carry the subscript \lq ext\rq, are not 
included in the model. Their concentrations have been assumed fixed so that
they are not dynamical variables. The same is true for $CO_2$, NADP, NADPH and 
starch. The solutions 
of biological interest are those which are positive (i.e. all concentrations 
are positive). Solutions which are non-negative but not positive are of 
interest as limits of the biologically applicable ones. Solutions which start 
positive remain positive and those which start non-negative remain 
non-negative. This is a consequence of Lemma 1 in
the next section. Let $S$ denote the positive orthant in the space of 
concentrations and $\bar S$ its closure. Then $S$ and $\bar S$ are invariant 
under the evolution. The total quantity of phosphate in the system is 
\begin{eqnarray}
&&c_P=2x_{RuBP}+x_{PGA}+2x_{DPGA}+x_{GAP}+x_{P_i}+x_{DHAP}\nonumber\\
&&+2x_{FBP}+x_{F6P}+x_{E4P}+x_{X5P}+2x_{SBP}+x_{S7P}\nonumber\\
&&+x_{R5P}+x_{Ru5P}+x_{G6P}+x_{G1P}+3x_{ATP}+2x_{ADP}.
\end{eqnarray} 
It is conserved for biological reasons and of course it follows directly from 
the evolution equations that the time derivative of this quantity is zero. In
particular it is a bounded function of time. Since all substances occurring 
in the system contain phosphate it follows that all concentrations are bounded 
and that solutions of the system exist globally in time.  

In the original Pettersson model a distinction is made between fast reversible 
and slow irreversible reactions when specifying the kinetics. The irreversible 
reactions are $RuBP\to PGA$, $FBP\to F6P$, $SBP\to S7P$, $Ru5P\to RuBP$, 
$ADP+P_i\to ATP$, $G1P+ATP\to ADP+P_i$, $PGA\to P_i$, $GAP\to P_i$ and 
$DPGA\to P_i$. In the original Poolman model these reactions are given the 
same kinetics as in the Pettersson model. These kinetics are chosen on the 
basis of experimental data. 
In the Poolman model the remaining reactions are given 
mass action kinetics. The evolution equations can be expressed in terms
of reaction rates $v_i$ without the kinetics being fixed. The result is
\begin{eqnarray}
&&\frac{d x_{RuBP}}{dt}=v_{13}-v_1,\\
&&\frac{d x_{PGA}}{dt}=2v_1-v_2-v_{\rm PGA},\\
&&\frac{d x_{DPGA}}{dt}=v_2-v_3,\\ 
&&\frac{d x_{ATP}}{dt}=v_{16}-v_2-v_{13}-v_{\rm st},\\
&&\frac{d x_{GAP}}{dt}=v_3-v_4-v_5-v_7-v_{10}-v_{\rm GAP},\\
&&\frac{d x_{DHAP}}{dt}=v_4-v_5-v_8-v_{\rm DHAP},\\
&&\frac{d x_{FBP}}{dt}=v_5-v_6,\\
&&\frac{d x_{F6P}}{dt}=v_6-v_7-v_{14},\\
&&\frac{d x_{E4P}}{dt}=v_7-v_8,\\
&&\frac{d x_{X5P}}{dt}=v_7+v_{10}-v_{12},\\
&&\frac{d x_{SBP}}{dt}=v_8-v_9,\\
&&\frac{d x_{S7P}}{dt}=v_9-v_{10},\\
&&\frac{d x_{R5P}}{dt}=v_{10}-v_{11},\\
&&\frac{d x_{Ru5P}}{dt}=v_{11}+v_{12}-v_{13},\\
&&\frac{d x_{G6P}}{dt}=v_{14}-v_{15},\\
&&\frac{d x_{G1P}}{dt}=v_{15}-v_{\rm st}+v_{17}\\
&&\frac{d x_{P_i}}{dt}=v_3+v_6+v_9+v_{PGA}+v_{GAP}+v_{DHAP}+2v_{\rm st}-v_{16}-v_{17}.
\end{eqnarray}
These equations are identical to those in \cite{pettersson88} except for the
fact that a rate $v_{17}$ has been added to accommodate the degradation of 
starch in the Poolman model and that the equation for the concentration of
inorganic phosphate has been included explicitly. In \cite{pettersson88}
this last equation was omitted since it can be computed from the other 
concentrations using the conservation law for the total amount of 
phosphate. In \cite{pettersson88} the slow reactions correspond to the $v_i$ 
with $i=1,6,9,13,16$ and the rates of the export reactions $v_{\rm PGA}$, 
$v_{\rm GAP}$ and $v_{\rm DHAP}$. Poolman also takes the additional reaction he
introduces to be a slow reaction, with rate \cite{poolman99}
\begin{equation}
v_{17}=\frac{V_{17}x_{P_i}}{x_{P_i}+K_m(1+K_{i17}^{-1}x_{G1P})}.
\end{equation}
The expressions for the $v_i$ in the 
Pettersson-MA and Poolman-MA models can be read off by 
comparing these equations with the evolution equations for those models given 
earlier. This gives the expressions for the $v_i$ in the Poolman model in the 
case of the fast reactions. The rates of the slow reactions in the Poolman
model apart from $v_{17}$ are taken from those in the Pettersson model and are 
as follows
\begin{eqnarray}
&&v_1=\frac{V_1x_{RuBP}}{x_{RuBP}+M_1},\\
&&v_6=\frac{V_6x_{FBP}}{x_{FBP}+K_{m6}(1+K^{-1}_{i61}x_{F6P}+K^{-1}_{i62}x_{P_i})},\\
&&v_9=\frac{V_6x_{SBP}}{x_{SBP}+K_{m9}(1+K^{-1}_{i9}x_{P_i})},\\
&&v_{13}=\frac{V_{13}x_{Ru5P}x_{ATP}}{M_{13}},\\
&&v_{16}=\frac{V_{16}x_{ADP}x_{P_i}}{(x_{ADP}+K_{m161})(x_{P_i}+K_{m162})},\\
&&v_{PGA}=\frac{V_{ex}x_{PGA}}{NK_{PGA}},\\
&&v_{GAP}=\frac{V_{ex}x_{GAP}}{NK_{GAP}},\\
&&v_{DHAP}=\frac{V_{ex}x_{DHAP}}{NK_{DHAP}},\\
&&v_{st}=\frac{V_{st}x_{G1P}x_{ATP}}{(x_{G1P}+K_{mst1})M_{st}}.
\end{eqnarray}
Here 
\begin{eqnarray}
&&M_1=K_{m1}\left(1+\frac{x_{PGA}}{K_{i11}}+\frac{x_{FBP}}{K_{i12}}
+\frac{x_{SBP}}{K_{i13}}+\frac{x_{P_i}}{K_{i14}}+\frac{x_{NADPH}}{K_{i14}}\right),\\
&&M_{13}=[x_{Ru5P}+K_{m131}(1+K^{-1}_{i131}x_{PGA}
+K^{-1}_{i132}x_{RuBP}+K^{-1}_{i132}x_{P_i})]\nonumber\\
&&\times [x_{ATP}(1+K^{-1}_{i134}x_{ADP})+K_{m132}(1+K^{-1}_{i135}x_{ADP})],\\
&&M_{st}=(1+K_{ist}^{-1}x_{ADP})
(x_{ATP}+K_{mst2}(1+K_{mst2}x_{P_i}\nonumber\\
&&\times (K_{ast1}x_{PGA}+K_{ast2}x_{F6P}+K_{ast3}x_{FBP})^{-1}),\\
&&N=1+(1+x_{P_{ext}}^{-1})\left(\frac{x_{P_i}}{K_{P_i}}
+\frac{x_{PGA}}{K_{PGA}}+\frac{x_{GAP}}{K_{GAP}}+\frac{x_{DHAP}}{K_{DHAP}}\right).
\end{eqnarray} 
It been pointed out in \cite{arnold14} that the expression for $v_{st}$ in
\cite{pettersson88} is incorrect and in the expression for $M_{st}$ given 
above we have used the replacement proposed in \cite{arnold14}.  
In the original Pettersson model the fast reactions are assumed to be at 
equilibrium which leads to a system of differential-algebraic equations.
The latter model will not be treated further in the present paper.

In the Calvin cycle most reactions conserve the number of carbon atoms.
There are, however, some inflow and outflow reactions for which the number of
carbon atoms which are in the chloroplast and not stored as starch is not
conserved. Hence the total number of carbon atoms in the substances included
in the model is not conserved. It is nevertheless useful to consider the
following small modification of the total number of carbon atoms. Define a 
quantity $L_1$ by
\begin{eqnarray}
&& 5L_1=5x_{RuBP}+\frac52x_{PGA}+3x_{DPGA}+3x_{GAP}+3x_{DHAP}+6x_{FBP}\nonumber\\
&&+6x_{F6P}+4x_{E4P}+5x_{X5P}+7x_{SBP}+7x_{S7P}+5x_{R5P}+5x_{Ru5P}\nonumber\\
&&+6x_{G6P}+6x_{G1P}.
\end{eqnarray} 
This is inspired by a Lyapunov function constructed by trial and error 
in \cite{rendall14} which is related in a similar way to the total number of
carbon atoms. Its time derivative is given by
\begin{eqnarray}
&&\frac{d}{dt}(5L_1)=\left(\frac12 k_3x_{ATP}-\frac52 k_{28}\right)x_{PGA}
-\frac12 k_2x_{DPGA}x_{ADP}\nonumber\\
&&-6k_{27}x_{ATP}x_{G1P}-3k_{29}x_{GAP}-3k_{30}x_{DHAP}+k_{32}x_{P_i}
\end{eqnarray}
for the Pettersson-MA and Poolman-MA models.

In the Pettersson-MA model if it is assumed that the parameters are such that
$k_3c_A\le 5k_{28}$ then $L_1$ is a Lyapunov function and it is strictly 
decreasing for positive solutions. It follows that in this case all 
$\omega$-limit points of a positive solution are on the boundary of $S$ and 
that, in particular, there are no positive stationary solutions. This parameter
restriction is the direct analogue of one found to play a role in the dynamics
of the simple model including ATP considered in Section 6 of \cite{rendall14}.  

\section{Potential $\omega$-limit points}

In this section information will be obtained on the location of $\omega$-limit
points of positive solutions of the Pettersson-MA and Poolman-MA systems. Many 
of the arguments use the following simple lemma.

\noindent
{\bf Lemma 1} Consider an ordinary differential equation of the form 
$\dot u(t)=-a(t)u(t)+b(t)$ where $a$ and $b$ are non-negative continuous 
functions and a solution $u(t)$ which satisfies $u(t_0)\ge 0$ for some $t_0$. 
Then if $b(t_1)>0$ for some $t_1>t_0$ it follows that $u(t_1)>0$.

\noindent
{\bf Proof} Suppose first that $u(t_0)>0$. Then $u$ remains positive for $t$ 
slightly larger than $t_0$ and as long as it does so the equation can be 
rewritten as $\frac{d}{dt}(\log u)=-a+\frac{b}{u}$. Thus 
$\frac{d}{dt}(\log u)\ge -a$ and $\log u$ cannot tend to $-\infty$ in finite 
time. Hence $u$ cannot tend to zero in finite time and $u$ remains strictly 
positive. It then follows using the continuous dependence of solutions on
initial data that if $u$ starts non-negative it stays non-negative. Now 
suppose there were a time $t_1>t_0$ with $u(t_1)=0$ and $b(t_1)>0$. Then 
$\dot u(t_1)>0$. This implies that $u(t)$ is negative for $t$ slightly less
than $t_1$, in contradiction to what has already been proved. This completes
the proof of the lemma.

It is also often useful to apply the contrapositive statement: if $u(t_1)=0$
then $b(t_1)=0$. In the applications of this lemma below $u$ will be one of the 
concentrations in a photosynthesis model and the functions $a$ and $b$ are 
obtained by setting the other concentrations to their values in a fixed 
non-negative solution.

The strategy is now to successively obtain restrictions on the position of an
$\omega$-limit point of a positive solution.

\noindent
{\bf Lemma 2} At an $\omega$-limit point of a positive solution of the 
Pettersson-MA or Poolman-MA model on the boundary of $S$ the 
concentrations of the following substances vanish: RuBP, PGA, DPGA, GAP, DHAP, 
FBP, SBP.

\noindent
{\bf Proof} The proof consists of repeated applications of Lemma 1 to the 
solution of the dynamical system passing through the $\omega$-limit point
being considered. Consider an $\omega$-limit point where $x_{GAP}>0$. Then the 
evolution equation for $x_{P_i}$ shows that this quantity is positive at the 
given point. In the same way the concentrations of the following quantities 
are positive: DHAP, FBP, F6P, E4P, X5P, SBP, S7P, R5P, Ru5P, G6P, G1P, ATP, 
RuBP, PGA, DPGA. This implies that the $\omega$-limit point is in the interior 
of $S$, contrary to the assumptions of the lemma. Thus it follows that in fact 
$x_{GAP}=0$. The evolution equation for $x_{GAP}$ then implies that the 
concentrations of DPGA, DHAP and FBP vanish. The evolution equation for 
$x_{DHAP}$ then implies that the concentration of SBP vanishes. To proceed 
further we need to distinguish between the cases where the concentration of 
ATP is non-zero or zero at the given point. In the first case we can conclude 
successively that the concentrations of PGA and RuBP vanish, completing the 
proof. In the second case $x_{P_i}=0$ and the evolution equation for $x_{P_i}$ 
implies that $x_{PGA}=0$. It then follows as in the first case that $x_{RuBP}=0$. 

\noindent
{\bf Lemma 3} At an $\omega$-limit point of a positive solution of the 
Pettersson-MA or Poolman-MA model on the boundary of $S$ the 
concentrations of $X5P$, $R5P$ and $Ru5P$ vanish.

\noindent
{\bf Proof} The evolution equations for $X5P$, $R5P$ and $Ru5P$ show that
the concentrations of all three substances vanish at an $\omega$-limit 
point of a stationary solution if and only if any one of them does.
However supposing that none of them vanishes leads to a contradiction in
the evolution equation for $GAP$.

%Note that when the conclusions of Lemma 2 and Lemma 3 hold a closed system of
%equations is obtained for the concentrations of ATP, $P_i$, F6P, G1P and 
%G6P. Thus for further analysis of $\omega$-limit points of the original 
%system it suffices to analyse $\omega$-limit points of the sub-system.

\noindent
{\bf Lemma 4} At an $\omega$-limit point on the boundary of $S$ of a positive 
solution of the Pettersson-MA model with $k_3c_A\le 5k_{28}$ either the 
concentrations of $G1P$, $G6P$ and $F6P$ vanish or all three are non-vanishing 
and $x_{ATP}=x_{P_i}=0$.

\noindent
{\bf Proof} In the Pettersson-MA and Poolman-MA models the evolution equations 
for $G1P$, $G6P$ and $F6P$ show that the concentrations of all three 
substances vanish at an $\omega$-limit point of a positive solution if and 
only if any one of them does. In the case of the Pettersson-MA model with 
$k_3c_A\le 5k_{28}$ it follows from the expression for the time derivative of 
$L_1$ that $x_{ATP}x_{G1P}$ vanishes at any $\omega$-limit point. If 
$x_{G1P}\ne 0$ at that point then $x_{ATP}=0$ there. It follows that $x_{P_i}=0$
at that point

\noindent
{\bf Lemma 5} At an $\omega$-limit point on the boundary of $S$ of a positive 
solution of the Poolman-MA model either the concentrations of all three hexoses
$G1P$, $G6P$ and $F6P$ vanish or none of them does so. One of the following 
three cases holds: 

\begin{enumerate}
\item $x_{P_i}=0$, all hexose concentrations are zero or 
\item $x_{P_i}=0$, $x_{ATP}=0$, all hexose concentrations are non-zero or 
\item $x_{P_i}\ne 0$, $x_{ATP}\ne 0$, all hexose concentrations are non-zero.
\end{enumerate}

\noindent
{\bf Proof} The first statement is part of Lemma 4. When $x_{G1P}=0$ at
an $\omega$-limit point on the boundary it follows from the evolution equation 
for $x_{G1P}$ that $x_{P_i}=0$. This is case 1. Otherwise all hexose
concentrations are non-zero. In that case if $x_{P_i}=0$ the evolution equation 
for $x_{P_i}$ shows that $x_{ATP}=0$. On the other hand if $x_{ATP}=0$ the 
evolution equation for $x_{ATP}=0$ shows that $x_{P_i}=0$. 

Any stationary solution on the boundary satisfies the conditions derived above
for $\omega$-limit points. In the Petterson-MA model a stationary solution
with non-vanishing hexose concentrations satisfies $x_{P_i}=0$ and $x_{ATP}=0$
without the restriction on the parameters occurring in Lemma 4. If, on the 
other hand, the concentrations of the hexoses vanish then the conditions for 
stationary solutions reduce to the condition that either $x_{P_i}=0$ or 
$x_{ADP}=0$. Given the fact that the concentrations of $x_{E4P}$ and $x_{S7P}$ 
can be prescribed freely we see that there are two three-parameter families 
of stationary solutions which meet at the point where both $x_{P_i}$ and 
$x_{ADP}$ are zero. In the case that the concentrations of the hexoses do not 
vanish we get a three-parameter family of stationary solutions satisfying the 
conditions $k_{23}x_{G6P}=k_{24}x_{F6P}$ and $k_{25}x_{G1P}=k_{26}x_{G6P}$. 
Consider now the Poolman-MA model. When $x_{P_i}=0$ the two systems agree and
so the set of stationary solutions is identical. It remains to consider the 
possibility that there are stationary solutions of the Poolman-MA model with
$x_{P_i}\ne 0$. They would belong to case 3. of Lemma 5 and satisfy the relation
$k_{31}(c_A-x_{ATP})=k_{32}$. Substituting this into the equation for $x_{P_i}$
gives a contradiction and so stationary solutions of this type do not exist. 

\section{Linearization about the $\omega$-limit points}

In Lemma 2 and Lemma 3 a set was identified, call it $Z$, where any 
$\omega$-limit point of a solution of the Petterson-MA model or the Poolman-MA 
model must lie. Consider now the linearization of the full model about a point 
of $Z$. The linearized quantity for a given substance is denoted by a $y$ with 
the name of that substrate as a subscript. The aim is to find a block upper
triangular form for the linearization, so as to obtain information about 
its eigenvalues. The linearized equation for FBP depends only on that 
substance and so can be split off, contributing a negative eigenvalue. 
At a general point of $Z$ it is difficult to proceed further with this
strategy. The simplest points about which to linearize are the points $z_0$
where all carbohydrate concentrations and $x_{P_i}$ are zero. Denote the
concentration of ATP at the point $z_0$ by $a$. The linearization is
\begin{eqnarray}
&&\frac{d y_{RuBP}}{dt}=-k_1 y_{RuBP}+k_{22}ay_{Ru5P},\\
&&\frac{d y_{PGA}}{dt}=2k_1 y_{RuBP}+k_2(c_A-a)y_{DPGA}-(k_3a+k_{28})y_{PGA},\\
&&\frac{d y_{DPGA}}{dt}=-[k_2(c_A-a)+k_5]y_{DPGA}+k_3ay_{PGA},\\
&&\frac{d y_{ATP}}{dt}=k_2(c_A-a)y_{DPGA}-k_3ay_{PGA}-k_{22}a y_{Ru5P}\nonumber\\ 
&&-k_{27}ay_{G1P}+k_{31}(c_A-a)y_{P_i},\\
&&\frac{d y_{GAP}}{dt}=k_5y_{DPGA}+k_6y_{DHAP}-k_7y_{GAP}
-k_{29}y_{GAP},\\
&&\frac{d y_{P_i}}{dt}=k_5y_{DPGA}+k_{15}y_{SBP}
+2k_{27}ay_{G1P}+k_{28}y_{PGA}\nonumber\\
&&+k_{29}y_{GAP}+k_{30}y_{DHAP}-k_{31}(c_A-a)y_{P_i}-k_{32}y_{P_i},\\
&&\frac{d y_{DHAP}}{dt}=-k_6y_{DHAP}+k_7y_{GAP}+k_{13}y_{SBP}
-k_{30}y_{DHAP},\\
&&\frac{d y_{F6P}}{dt}=k_{23}y_{G6P}
-k_{24}y_{F6P},\\
&&\frac{d y_{E4P}}{dt}=k_{13}y_{SBP},\\
&&\frac{d y_{X5P}}{dt}=k_{20}y_{Ru5P}-k_{21}y_{X5P},\\
&&\frac{d y_{SBP}}{dt}=-(k_{13}+k_{15})y_{SBP},\\
&&\frac{d y_{S7P}}{dt}=k_{15}y_{SBP},\\
&&\frac{d y_{R5P}}{dt}=k_{18}y_{Ru5P}-k_{19}y_{R5P},\\
&&\frac{d y_{Ru5P}}{dt}=-k_{18}y_{Ru5P}+k_{19}y_{R5P}-k_{20}y_{Ru5P}\nonumber\\
&&+k_{21}y_{X5P}-k_{22}ay_{Ru5P},\\
&&\frac{d y_{G6P}}{dt}=-k_{23}y_{G6P}+k_{24}y_{F6P}+k_{25}y_{G1P}-k_{26}y_{G6P},\\
&&\frac{d y_{G1P}}{dt}=-k_{25}y_{G1P}+k_{26}y_{G6P}-k_{27}ay_{G1P}
+k_{32}y_{P_i}
\end{eqnarray}
Here the quantity $y_{FBP}$ has been set to zero and its evolution equations
omitted since it plays no role in what follows.

\noindent
{\bf Lemma 6} Let $L$ be the linearization of the right hand side of the 
equations of Pettersson-MA model at a point of the form $z_0$. Then all 
eigenvalues of $L$ have non-positive real part and at most four have 
zero real part. The multiplicity of the eigenvalue zero is four when
$x_{ATP}=c_A$ and three otherwise. 

\noindent
{\bf Proof} The equation for SBP is decoupled from the others and contributes
a negative eigenvalue. The rows and columns corresponding to this variable 
can be discarded. The variables E4P and S7P can then be treated in the same
way, both contributing zero eigenvalues. Consider next the evolution equations 
for the variables $y_{G1P}$, $y_{G6P}$ and $y_{F6P}$. They form a closed system 
and define a submatrix which can be analysed on its own. The trace and 
determinant of this block are negative. The characteristic polynomial can 
easily be computed and the Routh-Hurwitz criterion \cite{gantmacher59} shows 
that all  
eigenvalues have negative real parts. The equations for $y_{X5P}$, $y_{Ru5P}$ 
and $y_{R5P}$ can be treated in exactly the same way. Once this has been
done the quantity $y_{RuBP}$ can be handled in the same way as the variables 
for FBP and SBP. The quantities $y_{PGA}$ and $y_{DPGA}$ define a 2 by 2 block 
which has negative trace and positive determinant. It contributes eigenvalues 
with negative real parts. Next $y_{GAP}$ and $y_{DHAP}$ can be treated together. 
They also contribute a 2 by 2 matrix with negative trace and positive 
determinant and therefore two eigenvalues with negative real parts. The 
equation for $y_{P_i}$ is now decoupled and gives the eigenvalue 
$-k_{31}(c_A-a)$, where $a$ is the concentration of ATP at the given point. 
Finally $y_{P_i}$ contributes a zero eigenvalue. This completes the proof.

\noindent
{\bf Lemma 7} Let $L$ be the linearization of the right hand side of the 
equations of the Poolman-MA model at the point $z_0$ with $x_{ADP}=0$. For 
$k_{32}$ sufficiently small there is one eigenvalue with positive real part, 
three zero eigenvalues and all the other eigenvalues have negative real part

\noindent
{\bf Proof} The methods in the proof of Lemma 6 can be used in a very similar
way to eliminate all the variables except $y_{G1P}$, $y_{G6P}$, $y_{F6P}$, $y_{ATP}$
and $y_{P_i}$ from consideration in the case $k_{32}\ne 0$. In the remaining
system of five equations $y_{ATP}$ does not occur on the right hand side and
so can also be eliminated, producing a further zero eigenvalue. To complete
the argument the eigenvalues of a four by four matrix must be examined. They 
depend continuously on the parameter $k_{32}$. When $k_{32}=0$ three of the 
eigenvalues have negative real parts according to Lemma 6. These eigenvalues 
retain this property for sufficiently small values of $k_{32}$. In this regime 
the sign of the remaining eigenvalue, which is real, is the opposite of that 
of the determinant. The determinant is $-k_{24}k_{26}k_{27}k_{32}c_A$, which is 
negative.

Lemma 6 and Lemma 7 can be used in combination with the reduction theorem
(see \cite{kuznetsov10}, Theorem 5.4) to obtain results on positive solutions 
of the nonlinear equations which start 
close to a point of the from $z_0$. Consider first the case of the 
Petterson-MA model with $x_{ATP}=c_A$. The centre manifold is four-dimensional
and is given by the vanishing of all variables except $x_{E4P}$, $x_{S7P}$,
$x_{ATP}$ and $x_{P_i}$. The restriction of the system to that manifold is the 
product of a trivial system for $x_{E4P}$ and $x_{S7P}$ with a two-dimensional 
system which is easily analysed. The conclusion is that solutions which start 
close to $z_0$ have the property that all concentrations converge to limits as 
$t\to\infty$. Generically the limits of $x_{E4P}$ and $x_{S7P}$ are strictly 
positive and precisely one of the limiting concentrations $x_{ADP}$ and 
$x_{P_i}$ is positive. If instead $x_{ATP}<c_A$ at $z_0$ then the centre manifold 
is three-dimensional. Again all concentrations tend to limits, with the limit 
of $x_{P_i}$ being zero.

In the case of the Poolman-MA model and the point of the form $z_0$ with
$x_{ADP}=0$ there is a one-dimensional unstable manifold and generic solutions 
starting close to $z_0$ do not stay close to $z_0$. Thus $z_0$ is stable for 
the Pettersson-MA model and passing to the Poolman-MA model destabilises it.
The production of sugar from starch in the Poolman-MA model means that 
there is no direct analogue of the argument which was used to show that 
the function $L_1$ is a Lyapunov function for the Pettersson-MA model
for certain values of the parameters. Instead it is possible to identify a 
parameter regime where $-L_1$ is a Lyapunov function.  

The total amount of phosphate can be written as the sum of the amount 
contained in the adenosine phosphates, the amount of inorganic phosphate
and the rest $c_P=c_A+P_R+x_{P_i}$. For a positive solution the
constant $c_1=c_P-c_A$ is positive. The quantity $P_R$ can be 
bounded in terms of $L_1$, counting the carbon atoms, with the result that 
$P_R\le 10L_1$. Hence $x_{P_i}\ge c_1-10L_1$. It follows that
\begin{equation}
\frac{dL_1}{dt}\ge 6k_{32}x_{P_i}-c_2L_1
\end{equation}
where $c_2$ is a constant which depends only on the reaction constants and 
the total amount of adenosine phosphates. Hence
\begin{equation}
\frac{dL_1}{dt}\ge 6k_{32}(c_1-10L_1)-c_2L_1
=6k_{32}[c_1-(10+c_2/(6k_{32}))L_1].
\end{equation}
This quantity is positive when $c_1-(10+c_2/(6k_{32})>0$. Let
$m=\frac{6k_{32}c_1}{60k_{32}+c_2}$. Then whenever $L_1$ is less than $m$ its
time derivative is positive. It follows that $L_1$ is eventually at least $m$.
In particular, it cannot tend to zero for $t\to\infty$. The latter fact 
follows from the computations of the previous section but here an 
additional quantitative lower bound is obtained for the concentrations
of the sugar phosphates at late times. It can be concluded that
\begin{equation}
\liminf_{t\to\infty} (4x_{E4P}+5x_{X5P}+6x_{G1P}+6x_{G6P}+6x_{F6P})\ge m.
\end{equation}

The main results of this section will be summed up in a theorem:

\noindent
{\bf Theorem 1} There is a non-empty open set of positive initial data for 
which the corresponding solutions of the Pettersson-MA model converge to the 
boundary of $S$ as $t\to\infty$. They include all data for which the 
concentrations of all carbohydrate phosphates and inorganic phosphate are 
bounded by a sufficiently small constant $\epsilon$. In the latter case the 
concentrations of all carbohydrate phosphates except E4P and S7P tend to zero 
while $x_{E4P}$ and $x_{S7P}$ remain small as $t\to\infty$. On the other hand, 
the solutions of the Poolman-MA model corresponding to these data do not have 
the property that the concentrations of all carbohydrate phosphates remain 
smaller than a constant $\delta>0$ for a suitable choice of $\epsilon$.

\section{The Poolman model}

The original model of Poolman uses kinetics which are not mass action. In this 
section we investigate which of the results for the Poolman-MA model have 
analogues in this case. The total amount of phosphate is conserved and this 
can be derived from the evolution equations independent of the kinetics 
given above together with the fact that $(d/dt)(x_{ADP})=-(d/dt)(x_{ATP})$.
Lemma 1 can be applied to this model in exactly the same way as it was 
applied to the Poolman-MA model. For all that is important in that context 
are the signs (positive, negative or zero) of the reaction rates when certain 
concentrations are zero or non-zero. These relations are not changed when 
the mass action kinetics are replaced by the more complicated kinetics in the 
original model. The modulation occurs because the reaction rates are 
multiplied by positive factors depending on the concentrations of other
substances. It can be further concluded that the analogues of Lemma 2 and 
Lemma 3 hold for the Poolman system. The time derivative of $L_1$ is given by
\begin{equation}
\frac{d}{dt}(5L_1)=\frac12 v_2-\frac52 v_{PGA}-3v_{GAP}-3v_{DHAP}-6v_{\rm st}+6v_{17}.
\end{equation}
The manifestly positive term on the right hand side of this equation is given
by $6v_{17}=6k_{32}x_{P_i}$. In order to show that the estimate obtained in the 
case of the Poolman-MA model extends to the Poolman model it is enough to 
obtain lower bounds for all the other terms on the right hand side of the
equations by negative multiples of $L_1$. The quantity $v_2$ is a combination
of two mass action terms and there is nothing new to do. The remaining terms
admit the desired type of bound.  

\section{Conclusions and outlook}

In this paper information has been obtained about the dynamics of some models 
for the Calvin cycle of photosynthesis. It was shown for the Pettersson-MA model
that if the reaction constant determining the rate of conversion of PGA to 
DPGA is small enough compared to that determining the rate of export of PGA 
from the chloroplast then the concentration
of some sugar phosphate must attain arbitarily small values at late times.
In particular for these parameter values the model admits no positive 
stationary solutions. This is a manifestation of the phenomenon of overload 
breakdown. It was further shown that under these circumstances all sugar 
phosphate concentrations except those of E4P, S7P, G1P, G6P and F6P tend to 
zero for $t\to\infty$. For initial data where the initial concentrations of 
$x_{P_i}$, $x_{E4P}$ and $x_{S7P}$ are sufficiently small all concentrations
tend to limits at late times and for generic data of this type the limits of 
the concentrations of E4P and S7P are not zero. In contrast it was shown that 
for the Poolman-MA model for generic data with $x_{P_i}$, $x_{E4P}$, $x_{S7P}$ 
and $x_{ATP}$ small it is not the case that these quantities stay small.
Thus, in accordance with the original motivation, the regime of overload 
breakdown is destabilized by the additional term introduced by Poolman.
This instability is also present in the original Poolman model.

It would be possible to introduce a hybrid model with the network of 
Poolman and the kinetics as in the Pettersson model. It might be possible
to obtain the Pettersson model as a singular limit of this hybrid model
but the details of how to do this in a mathematically rigorous way remain
to be worked out. There are a number of other interesting open questions 
concerning these models. Does the Poolman model (or the Poolman-MA model)
have a positive stationary solution for some values of the parameters? 
Numerical simulations in \cite{poolman99} indicate that there can be two
stable positive stationary solutions which exhibit hysteresis when the 
parameter $V_{16}$, which here represents the intensity of light, is varied.
Compare also the discussion in \cite{poolman00}. 
Does the Pettersson model (or the Pettersson-MA model) have a positive 
stationary solution for some values of the parameters? Numerical simulations in 
\cite{pettersson88} indicate that this is the case and that there can be one 
stable and one unstable solution. There are, however, no mathematical proofs 
that these features occur. To obtain a positive answer to one of these 
questions concerning the existence of stationary solutions it would suffice to 
show that in the case being considered positive solutions can have no 
$\omega$-limit points on the boundary of $S$ and for that it might suffice to 
linearize about stationary points on the boundary more general than those 
handled in this paper. Beyond the question of the existence of positive 
stationary solutions there are the questions of their multiplicity and 
stability. The analogous questions for the simple model including ATP 
considered in
\cite{grimbs11} and \cite{rendall14} are partly open. There are also many 
other models for the Calvin cycle in the literature and it would be desirable 
to do a careful mathematical study of their solutions. This is particularly 
relevant since several errors in the literature have been discovered 
which had gone uncorrected for many years \cite{jablonsky11}, \cite{arnold14}.

\end{document}